\documentclass[prd,twocolumn,superscriptaddress,showpacs]{revtex4}% Physical Review D

\usepackage{graphicx}% Include figure files
\usepackage{dcolumn}% Align table columns on decimal point
\usepackage{bm}% bold math

\newcommand{\met}{E_T\hspace{-1.1em}/\hspace{0.75em}}

\begin{document}

%\preprint{APS/123-QED}

%\bibliographystyle{revtex}

\title[lala]{Direct photon cross section with conversions at CDF}

\begin{abstract}

\font\eightit=cmti8
\def\r#1{\ignorespaces $^{#1}$}
\vskip0.25in
%\hfilneg
\begin{sloppypar}
\noindent
D.~Acosta,\r {14} T.~Affolder,\r 7 M.G.~Albrow,\r {13} D.~Ambrose,\r {36}   
D.~Amidei,\r {27} K.~Anikeev,\r {26} J.~Antos,\r 1 
G.~Apollinari,\r {13} T.~Arisawa,\r {50} A.~Artikov,\r {11}
% ...
\end{sloppypar}
\vskip .026in
\begin{center}
(CDF Collaboration)
\end{center}

\vskip .026in
\begin{center}
\r 1  {\eightit Institute of Physics, Academia Sinica, Taipei, Taiwan 11529, 
Republic of China} \\
\r 2  {\eightit Argonne National Laboratory, Argonne, Illinois 60439} \\
\r 3  {\eightit Istituto Nazionale di Fisica Nucleare, University of Bologna,
I-40127 Bologna, Italy} \\
\r 4  {\eightit Brandeis University, Waltham, Massachusetts 02254} \\
\r 5  {\eightit University of California at Davis, Davis, California  95616} \\
\r 6  {\eightit University of California at Los Angeles, Los 
Angeles, California  90024} \\ 
\r 7  {\eightit University of California at Santa Barbara, Santa Barbara, California 
93106} \\ 
\r 8 {\eightit Instituto de Fisica de Cantabria, CSIC-University of Cantabria, 
39005 Santander, Spain} \\
\r 9  {\eightit Carnegie Mellon University, Pittsburgh, Pennsylvania  15213} \\
\r {10} {\eightit Enrico Fermi Institute, University of Chicago, Chicago, 
Illinois 60637} \\
\r {11}  {\eightit Joint Institute for Nuclear Research, RU-141980 Dubna, Russia}
\\
\r {12} {\eightit Duke University, Durham, North Carolina  27708} \\
\r {13} {\eightit Fermi National Accelerator Laboratory, Batavia, Illinois 
60510} \\
\r {14} {\eightit University of Florida, Gainesville, Florida  32611} \\
\r {15} {\eightit Laboratori Nazionali di Frascati, Istituto Nazionale di Fisica
               Nucleare, I-00044 Frascati, Italy} \\
\r {16} {\eightit University of Geneva, CH-1211 Geneva 4, Switzerland} \\
\r {17} {\eightit Glasgow University, Glasgow G12 8QQ, United Kingdom}\\
\r {18} {\eightit Harvard University, Cambridge, Massachusetts 02138} \\
\r {19} {\eightit Hiroshima University, Higashi-Hiroshima 724, Japan} \\
\r {20} {\eightit University of Illinois, Urbana, Illinois 61801} \\
\r {21} {\eightit The Johns Hopkins University, Baltimore, Maryland 21218} \\
\r {22} {\eightit Institut f\"{u}r Experimentelle Kernphysik, 
Universit\"{a}t Karlsruhe, 76128 Karlsruhe, Germany} \\
\r {23} {\eightit Center for High Energy Physics: Kyungpook National
University, Taegu 702-701; Seoul National University, Seoul 151-742; and
SungKyunKwan University, Suwon 440-746; Korea} \\
\r {24} {\eightit Ernest Orlando Lawrence Berkeley National Laboratory, 
Berkeley, California 94720} \\
\r {25} {\eightit University College London, London WC1E 6BT, United Kingdom} \\
\r {26} {\eightit Massachusetts Institute of Technology, Cambridge,
Massachusetts  02139} \\   
\r {27} {\eightit University of Michigan, Ann Arbor, Michigan 48109} \\
\r {28} {\eightit Michigan State University, East Lansing, Michigan  48824} \\
\r {29} {\eightit Institution for Theoretical and Experimental Physics, ITEP,
Moscow 117259, Russia} \\
\r {30} {\eightit University of New Mexico, Albuquerque, New Mexico 87131} \\
\r {31} {\eightit Northwestern University, Evanston, Illinois  60208} \\
\r {32} {\eightit The Ohio State University, Columbus, Ohio  43210} \\
\r {33} {\eightit Osaka City University, Osaka 588, Japan} \\
\r {34} {\eightit University of Oxford, Oxford OX1 3RH, United Kingdom} \\
\r {35} {\eightit Universita di Padova, Istituto Nazionale di Fisica 
          Nucleare, Sezione di Padova, I-35131 Padova, Italy} \\
\r {36} {\eightit University of Pennsylvania, Philadelphia, 
        Pennsylvania 19104} \\   
\r {37} {\eightit Istituto Nazionale di Fisica Nucleare, University and Scuola
               Normale Superiore of Pisa, I-56100 Pisa, Italy} \\
\r {38} {\eightit University of Pittsburgh, Pittsburgh, Pennsylvania 15260} \\
\r {39} {\eightit Purdue University, West Lafayette, Indiana 47907} \\
\r {40} {\eightit University of Rochester, Rochester, New York 14627} \\
\r {41} {\eightit Rockefeller University, New York, New York 10021} \\
\r {42} {\eightit Instituto Nazionale de Fisica Nucleare, Sezione di Roma,
University di Roma I, ``La Sapienza," I-00185 Roma, Italy}\\
\r {43} {\eightit Rutgers University, Piscataway, New Jersey 08855} \\
\r {44} {\eightit Texas A\&M University, College Station, Texas 77843} \\
\r {45} {\eightit Texas Tech University, Lubbock, Texas 79409} \\
\r {46} {\eightit Institute of Particle Physics, University of Toronto, Toronto
M5S 1A7, Canada} \\
\r {47} {\eightit Istituto Nazionale di Fisica Nucleare, University of Trieste/\
Udine, Italy} \\
\r {48} {\eightit University of Tsukuba, Tsukuba, Ibaraki 305, Japan} \\
\r {49} {\eightit Tufts University, Medford, Massachusetts 02155} \\
\r {50} {\eightit Waseda University, Tokyo 169, Japan} \\
\r {51} {\eightit University of Wisconsin, Madison, Wisconsin 53706} \\
\r {52} {\eightit Yale University, New Haven, Connecticut 06520} \\
\end{center}
\vskip0.25in

We present a measurement of the isolated direct photon cross
section in $p\bar{p}$ collisions at $\sqrt{s} = 1.8$ TeV and
$|\eta| < 0.9$ using 
data collected between 1994 and 1995 by the Collider Detector
at Fermilab (CDF). The measurement is based on events where the
photon converts into an electron-positron pair in the material of the
inner detector, resulting in a two track event signature. To remove
$\pi^0 \rightarrow \gamma \gamma$ and $\eta \rightarrow \gamma
\gamma$ events from the data we use a new background
subtraction technique which takes advantage of the tracking 
information available in a photon conversion event. We find that the
shape of the cross section as a function of photon $p_T$ is
poorly described by next-to-leading-order QCD  predictions,
but agrees with previous CDF measurements. 

\end{abstract}

% PACS, the Physics and Astronomy
                             % Classification Scheme.

\date{\today}

\maketitle

%\section{\label{sec:level1}Frst-level heading:\protect\\ The line
%break was forced \lowercase{via} \textbackslash\textbackslash}

\section{Introduction}

The CDF Collaboration recently published a measurement
of the the direct photon cross section\cite{Acosta:2002ya}. 
This analysis found that the shape of the cross section as a 
function of $p_T$ is poorly described by next-to-leading-order 
(NLO) QCD calculations \cite{Gluck:1994iz}, and that 
the discrepancy persists at $\sqrt{s} = 1800$ GeV and 630 GeV. 
This conclusion is supported by measurements by
the D0 collaboration \cite{Abbott:1999kd}, \cite{Abazov:2001af}, 
and by other hadron-hadron experiments \cite{Apanasevich:1998hm}.

Photon measurements in hadron collisions are complicated by the
large number of $\pi^0 \rightarrow \gamma \gamma$ and 
$\eta \rightarrow \gamma \gamma$ events produced in these experiments.
These backgrounds are traditionally suppressed
by requiring that the photon be isolated from other energy 
in the calorimeter, 
but this requirement also eliminates some of the direct photon signal.
Special calculations which take the isolation requirement into account
have been developed in order to compare these measurements to 
NLO QCD \cite{Gluck:1994iz}.

To remove the remaining meson events from the data sample, 
experimentalists have relied upon understanding 
the shape and development of 
electromagnetic (EM) showers in the calorimeter . 
At CDF two techniques are used: a shower transverse profile 
method, and a pre-shower method \cite{Acosta:2002ya}. 
The datasets are
based on photon triggers, where a high $E_T$ EM  
shower is found in the central calorimeter with no associated
charged tracks. 

In this article we report on a new measurement of the direct 
photon cross section at CDF based on events
where the photon converts to an $e^+e^-$ pair in the detector 
material prior to passing through the central tracking chamber.
The EM showers in these events have
tracks associated with them, and so
are explicitly rejected by conventional photon measurements. 
Furthermore, the addition of tracking
information to the event makes possible a new background subtraction
technique which is systematically independent from the standard
calorimeter methods. 

The primary motivation for studying the direct photon
cross section is the potential to extract information about the 
parton distribution function (PDF) of the gluon inside the proton, 
due to the large contribution of $g q \rightarrow \gamma q$ 
diagrams to the process \cite{Vogelsang:1995bg}. This program has been 
frustrated by differences between the measurements
and calculations which are difficult to explain by altering
the gluon PDF alone \cite{Huston:1995vb},  
\cite{Huston:1998jj},  \cite{Apanasevich:1998ki}.
The direct photon cross section measurement with conversions
therefore serves as a cross check of conventional photon 
techniques, as well as a demonstration of a new method for future 
high $p_T$ photon studies.

\section{Detector and datasets}

The data was collected at the Fermilab TeVatron collider
between 1994 and 1995 (Run 1b) with a center-of-mass energy of 1.8 TeV. 
A detailed description of CDF in Run 1 may be
found elsewhere \cite{Abe:1988me}. Here we briefly describe those detector 
components critical for the conversion measurement. The central tracking
system consists of a silicon vertex detector (SVX), a vertex
TPC (VTX), and a large central tracking chamber (CTC). These
detectors are located inside a 1.4 Tesla solenoidal magnet.
The transverse momenta of charged particles in the tracking
system are measured primarily by the CTC, which has a momentum
resolution of $\sigma(p_T)/p^{2}_{T} = 0.002$ GeV$^{-1}$. 
Outside the tracking system are the CDF calorimeters, which
are subdivided in $\eta$\footnote{  Pseudorapidity ($\eta$) is
defined by $\eta = - \ln \tan (\theta/2)$, where $\theta$ is
the polar angle measured from the beamline.  }
 and $\phi$ into projective towers 
which point to the nominal $p\bar{p}$ interaction point at the
center of the detector.
The central region ($|\eta| < 1.1$) is instrumented with the central
electromagnetic (CEM), central hadronic (CHA), and wall hadronic 
(WHA) calorimeters. 
EM showers in the CEM generally deposit their energy in two or
three towers in $\eta$, and these towers are referred to as a CEM cluster.
The energy resolution of the CEM is 

\begin{displaymath}
\frac{\sigma(E)}{E}  =
\sqrt{\left ( \frac{13.5\%}{\sqrt{E \sin \theta}} \right )^2 + (1.6\%)^2}
\end{displaymath} 
where $\theta$ is the polar angle of the shower measured with respect
to the proton beam direction.
The CEM is equipped with a layer of crossed wire and strip gas chambers
(CES) located at a depth of six radiation lengths (the typical shower
maximum) to measure the transverse shape of the shower. 
A second layer of wire chambers, known as the CPR, is located between 
the solenoid and the CEM. The CPR is used as a pre-shower detector
in conventional photon measurements, with the 1.1 $X_0$ radiation 
lengths of the solenoid acting as the converting material. 
 
We use a three level trigger system to collect the two datasets used
in the photon cross section measurement with conversions. 
The first data sample, known as 
the 8 GeV electron data, requires a cluster in the CEM of at
least 8 GeV at Level 1. Level 2 requires an associated track
found by the fast hardware track finder (CFT) with $p_T >  7.5$ GeV, 
and an associated CES cluster found by a hardware cluster finder (XCES).
This trigger applies several electron identification requirements at
Level 3, including requirements on the transverse shape of the shower seen in 
the CES, the geometric matching between the shower and the track, the 
lateral sharing of the shower energy over the several CEM towers, and the
electromagnetic fraction of the shower. The integrated luminosity
of this dataset is 73.6 pb$^{-1}$.  

The second data sample, known as the 23 GeV photon data, requires
an 8 GeV CEM cluster at Level 1, but at Level 2 this requirement is 
increased to 23 GeV. 
The Level 2 trigger also applies an isolation requirement to the CEM cluster
by requiring that the neighboring calorimeter towers have $E_T < 4$ GeV. 
The 23 GeV photon trigger does not require that a track be found by 
the CFT, and it does not apply any electron identification requirements,
although at least one CES cluster must be found with more than 0.5 GeV
of energy at Level 3. Note that this trigger was designed to
collect non-conversion photons (hence its name), 
but since it does not 
veto photon candidates which have associated tracks, we can use
it to search for conversion events as well. The integrated 
luminosity is 83.7 pb$^{-1}$.

Inner detector photon conversions are characterized by  
two opposite sign CTC tracks which pass near each other
in the material of the beampipe, SVX, VTX, or inner cylinder of the
CTC . Two conversion identification 
requirements are applied to the raw CTC tracks. The
first requires that the absolute value of the 
difference between the track $\cot \theta$s
be less than 0.05.
The second requires that the 
absolute value of the distance between the tracks in the x-y plane
at the radial location where they are parallel be less than 0.3 cm. 
At least one of these tracks is required to point
at a CEM cluster, and the softer track is required to have 
$p_T > 0.4$ GeV. 

\begin{figure}
\scalebox{0.6}{\includegraphics{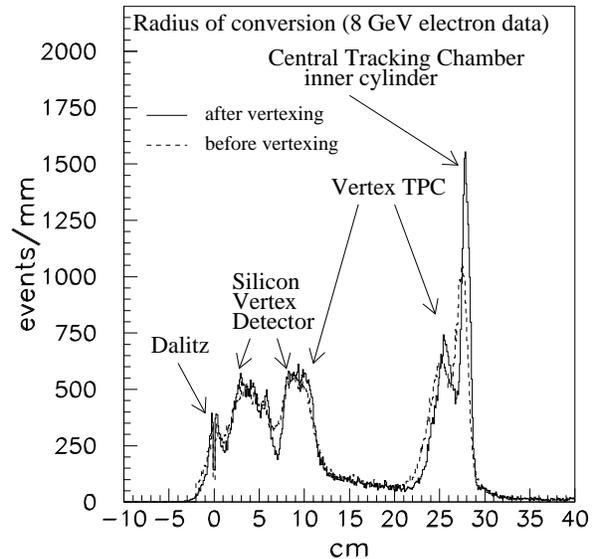}}
\caption
{Vertexed and un-vertexed radius of conversion distribution in
8 GeV electron data. The
peak at $r = 0$ labelled 'Dalitz' is due to $\pi^0 \rightarrow
e^+e^-\gamma$ decays and $\gamma^{*} \rightarrow e^+e^-$ events.
The 2 cm $ < r < $ 30 cm data selection requirement has
been released in this plot.}
\label{fig:xray}
\end{figure}

Track pairs satisfying these requirements are fitted to a 
conversion vertex. The fit requires that the tracks meet at a point in 
space where they are parallel, which improves the spatial and 
momentum resolutions of the reconstructed photon candidate. 
In addition, the vertex fit
partially corrects for a $p_T$ bias present in the raw conversion
tracks. This bias occurs when the spatial separation of the 
conversion tracks in the inner CTC superlayers is less than the 
two track resolution of the device.
The final requirement of the conversion selection is that the fitted 
conversion radius is required to be between 2 and 30 cm.
The radius of conversion distribution of the 8 GeV electron data
is shown in Figure \ref{fig:xray}.

%%%%%%%%%%%%%%%%%%%%%%%%%%%%%%%%%%%%%%%%%%%%%%%%%%%%%%%%%%%%%%%%%
%%%%%%%%%%%%%%%%%%%%%%%%%%%%%%%%%%%%%%%%%%%%%%%%%%%%%%%%%%%%%%%%

Conversion candidate
events at CDF divide naturally into two sets based on their 
detector topology. In 1-tower conversions both tracks point
to the same CEM tower, and in 2-tower conversions the tracks point 
to separate towers. 1-tower conversions have the potential 
to confuse the electron identification requirements applied by the 8 
GeV electron trigger at Level 3, due to the presence of two EM showers 
in the same CES chamber. Therefore in the 8 GeV electron data we require 
that the conversion be 2-tower. Conversely, in the 23 GeV photon
data we require that the conversion be 1-tower, in order to 
insure that the two datasets have no events in common. 

In a 1-tower event, the CEM cluster measures the summed $E_T$
of both tracks. In a 2-tower event, however, the softer track 
is outside the high $E_T$ cluster formed by the first track, and
its own $E_T$ usually falls below the clustering threshold of the 
CEM reconstruction algorithm. In this case, only the higher $E_T$ 
cluster is found, but the $p_T$ of both tracks is measured by the
CTC. When dividing the conversion events into $p_T$ bins, we
use the summed $E_T$ measurement for 1-tower events, while for 
2-tower events we use the $E_T$ measurement of the higher energy 
track plus the $p_T$ measurement of the lower energy track.

The offline data reduction proceeds as follows. For the 8 GeV electron
data we require a 2-tower conversion at $|\eta| < 0.9$ and 
$|z_0| < 60$ cm, where $z_0$ is the position of the primary event
vertex along the beamline measured from the center of the detector. 
The conversion must be associated 
with a fiducial CEM cluster, and we re-apply the 
electron identification requirements imposed by the Level 3 trigger.
The reconstructed CEM cluster must have $E_T > 8.0$ GeV, and 
at least one of the conversion track must have $p_T > 6.0$ GeV.
To suppress the contribution of $\pi^0$ and $\eta$ events we make two
isolation requirements. The first requires that the amount of energy found
in a cone of radius $ R = \sqrt{(\Delta \eta)^2 + (\Delta \phi)^2}$ 
less than 0.4 centered on the highest $E_T$ shower
be less than 1 GeV, excluding the energy in the CEM cluster itself. 
The second requires that no
extraneous tracks with $p_T > 0.4$ GeV point to the CEM cluster. Finally 
we require that the missing energy ($\met$) in this dataset be less than 25 
GeV in order to suppress a background due to $W \rightarrow e \nu$ events. 

For the 23 GeV photon data we require a 1-tower conversion at $|\eta| < 0.9$
and $|z_0| < 60$ cm
with a fiducial CEM cluster and the same isolation requirements. 
The CEM cluster must have $E_T > 28$ GeV and the conversion must 
have at least one track with $p_T > 8.0$ GeV. This 
dataset has no electron identification requirements, and no $\met$ requirement. 

There is one complication to the cone 0.4 isolation requirement in the
case of a 2-tower conversion. If the soft conversion track lands
outside the CEM cluster, but within the 0.4 cone, then
the cone energy sum is artificially enhanced by the energy of this track.
To remove this energy
the tower hit by the soft track and its closest neighbor in $\eta$
are excluded from the cone energy sum. In this case the area of the 
cone is slightly reduced, and to account for this the energy requirement is
reduced from 1.0 GeV to 0.87 GeV. This occurs in about 2/3 of all 
2-tower events. 

\section{Background subtraction}
\label{sec:back_sub}

\subsection{$\pi^0$ and $\eta$ backgrounds}

Most $\pi^0 \rightarrow \gamma \gamma $ 
and $\eta \rightarrow \gamma \gamma $    events are rejected by the 
isolation requirements. Those that remain are 
statistically subtracted from the data by a new technique based on 
$E/p$. $E/p$ is the ratio of the $E_T$ measured
in the CEM and the $p_T$ measured by the CTC. For a 1-tower
conversion the $E_T$ is the two-track summed energy measured by the CEM, 
and the $p_T$ is the sum of the two vertexed track momenta.
For a 2-tower conversion, the CEM cluster measures only the $E_T$ of the
higher energy track, and in this case the $E/p$ ratio includes
only the $E_T$ and vertexed $p_T$ of that track. 

Under this definition, the $E/p$ distribution
for a direct photon conversion 
should be a narrow peak centered on 1.0 
whose width is determined by the CTC and CEM resolutions. 
In a  $\pi^0$ or $\eta$ event, however, the
second unconverted photon usually showers in the same CEM cluster
as the high $E_T$ conversion electron. 
Therefore in a meson event the $E_T$ measures the $\pi^0$
energy, and the $p_T$ measures the energy of one of the decay
photons. Since two-body decay kinematics are understood, 
the shape of the meson $E/p$ distribution is relatively easy
to calculate with a Monte Carlo simulation.

To predict the signal $E/p$
distribution we generate direct photon events using PYTHIA 
version 6.115 \cite{Sjostrand:2000wi}. The prompt photon is tracked
through a material map of the CDF inner detector where
it is allowed to convert into an electron-positron pair.
The two tracks pass through the remaining material, 
where they are allowed to undergo bremsstrahlung, and 
through the tracking chamber and calorimeter. To simulate 
the $p_T$ and $E_T$ measurements the true track parameters 
are smeared by the known resolutions of the CTC and CEM.

%%%%%%%%%%%%%%%%%%%%%%%%%%%%%%%%%%%%%%%%%%%%%%%%%%%%%%%%%%%%%%%%%
\begin{figure}
\scalebox{0.6}{\includegraphics{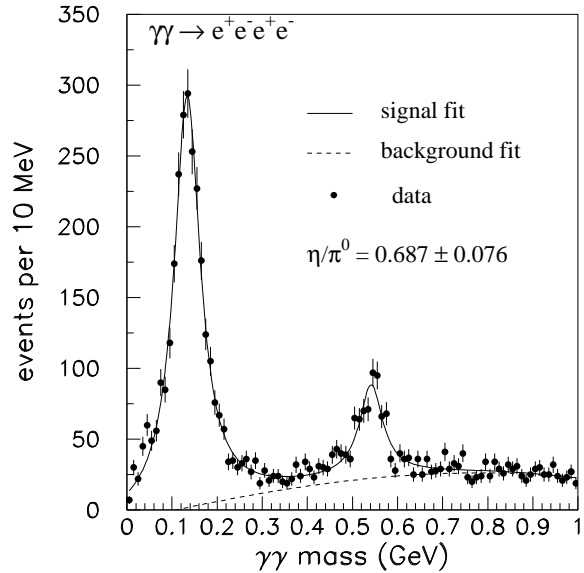}}
\caption{The diphoton mass spectrum of double conversion events in the data.
The data is fit to two Lorentzians plus a third order polynomial.
The polynomial is shown as the dotted line.
The $\pi^0$ and $\eta$ peaks are visible at 0.135 GeV and 0.547 GeV
respectively. The ratio of the areas of the two peaks, along
with the Monte Carlo prediction for the ratio of acceptances,
gives an $\eta / \pi^0$ production ratio of $0.687 \pm 0.076$.}
\label{fig:eta_pi_ratio}
\end{figure}
%%%%%%%%%%%%%%%%%%%%%%%%%%%%%%%%%%%%%%%%%%%%%%%%%%%%%%%%%%%%%%%%

We find that for the purpose
of predicting the meson $E/p$ distribution, it is adequate to 
simulate single mesons, rather that complete events, because
the mesons in the data are highly isolated.
The generated mesons decay to two photons which are tracked through
the detector in the same manner as the PYTHIA
direct photons. The two meson samples are 
combined using a $\eta / \pi^0$ production ratio of $0.69 \pm 0.08$, 
which we measured in the data using a sample of double conversion
($\pi^0 / \eta \rightarrow \gamma \gamma \rightarrow e^+e^-e^+e^-$)
events, as shown in Figure \ref{fig:eta_pi_ratio}. 
Fortunately, the $E/p$ distributions of $\pi^0$ and
$\eta$ are very similar (due to similar decay kinematics), 
so the production ratio used in the Monte Carlo has little effect
on their combined $E/p$ shape. We also use the double conversion
events to measure the meson $p_T$ spectrum to be used in the Monte Carlo.
We find that a power law with an exponent of negative six 
gives a good description of the data.

\begin{figure*}
\scalebox{0.55}{\includegraphics{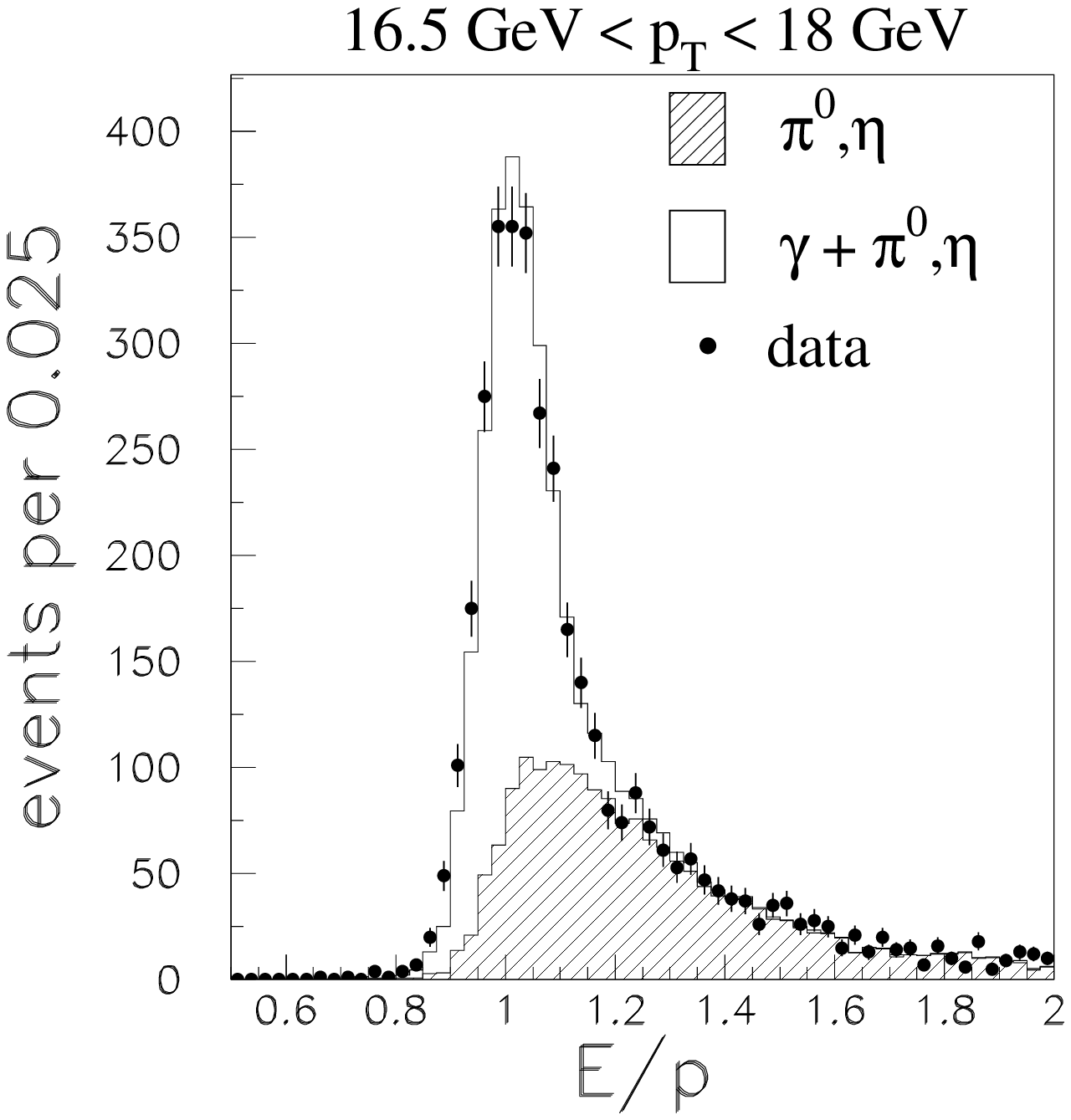}}
\scalebox{0.55}{\includegraphics{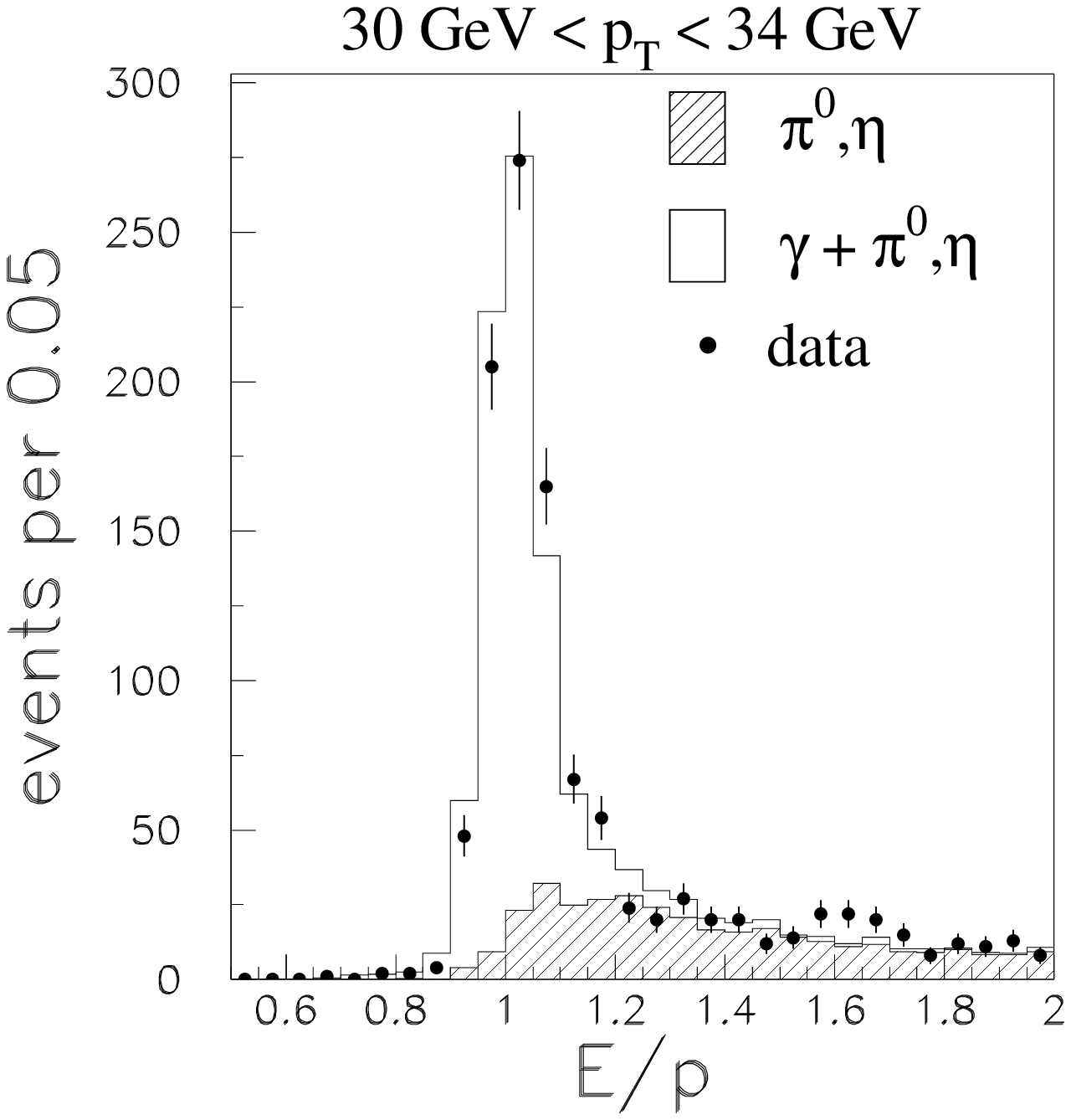}}
\caption{An example of the $E/p$ background subtraction 
fit in two $p_T$ bins. Left: the 16.5 to 18 GeV bin from the 8 GeV
electron (2-tower) data. Right: the 30 to 34 GeV bin from the
23 GeV photon (1-tower) data. }
\label{fig:prd_eop_plot}
\end{figure*}

We extract the number of signal candidates in each $p_T$
bin by performing a $\chi^2$ fit of the $E/p$ distributions observed in 
the data to the Monte Carlo signal and background templates. 
In the fit only the normalizations of the signal and background
are allowed to float.
Examples of two fits are shown in Figure 
\ref{fig:prd_eop_plot}, and the number of signal candidates
found in each $p_T$ bin is listed in Table 
\ref{tab:cross_section_summary}.

As seen in Figure \ref{fig:prd_eop_plot}, 
the narrow signal peak is quite distinct from the 
broad background distribution, and in general the fits to the data
are reasonable. In some fits, however, the signal
peak is shifted slightly with respect to the Monte Carlo 
prediction. This effect is due to a $p_T$ bias associated with
conversion tracking which occurs when the spatial separation
of the two tracks in the inner layers of the CTC is below the hit 
resolution. A hit level simulation of the tracking system reproduces this 
effect, but it is not simulated  by our fast Monte Carlo, so the 
templates do not reproduce this. Studies show that the $E/p$ shift is
no larger than 1\% in the 8 GeV electron data, and 2\% in 
the 23 GeV photon data \cite{Hall:2002sg}.

To determine a systematic uncertainty on the 
background subtraction due to this effect, we multiply the $E/p$
of each event in the 8 GeV electron data by a scale factors of 1.01
and 0.99. We then perform the 
$\chi^2$ fit again, and we take the change in the number of
signal candidates as a systematic uncertainty. Similarly we use
scale factors of 1.02 and 0.98 in the 23 GeV data to determine the
uncertainty.
For the 8 GeV electron data this error is 
$+12/-10\%$ at 10 GeV and decreases to less than $\pm 5\%$
above 20 GeV, while the error is less than $\pm 5\%$ for the 
23 GeV photon data.

%%%%%%%%%%%%%%%%%%%%%%%%%%%%%%%%%%%%%%%%%%%%%%%%%%%%%%%%%%%%%%%%%
%%%%%%%%%%%%%%%%%%%%%%%%%%%%%%%%%%%%%%%%%%%%%%%%%%%%%%%%%%%%%%%%

\subsection{Other backgrounds}

We consider two other potential sources of background. The
first is fake conversions, where two random tracks satisfy
the conversion identification requirements. In this case the soft 
conversion track is likely to be a hadron. A study
of the $E/p$ of the soft conversion tracks finds no evidence
for hadronic contamination, so we neglect this background.

A second source of background is due to high $p_T$ prompt
electrons, such as those produced in $W \rightarrow e \nu$ events.
These electrons often have a co-linear bremsstrahlung photon, 
and this photon may convert in the detector material and produce one
or two soft tracks. The soft tracks can form a high $p_T$ 
conversion candidate when combined with the prompt electron.
This background is the motivation for the $\met$  requirement
applied to the 8 GeV electron data, which would otherwise have 
significant $W \rightarrow e \nu$ contamination above 25 GeV.
In the 23 GeV photon data this background is less significant 
because these events are unlikely to satisfy the 1-tower topology.

To account for remaining prompt electron backgrounds in both
datasets, including any remaining W electrons, we have searched 
for hits in the SVX and VTX detectors
in events where the conversion occurs outside these detectors. 
These studies have indicated that in the 8 GeV electron data
there is no significant prompt electron contamination below 25 GeV,
and above 25 GeV we adopt a one sided 10\% systematic uncertainty.
In the 23 GeV photon data we adopt a one sided 3\% systematic
uncertainty in all $p_T$ bins. 

\section{Acceptance and efficiency}

The acceptance is evaluated with the
PYTHIA direct photon Monte Carlo, and includes the 
fiducial requirements, the 1-tower and 2-tower 
topological requirements, and the $E_T$ and $p_T$ requirements 
on the CEM clusters and tracks. For the 2-tower data (8 GeV
electron trigger), the acceptance is 33\% at 10 GeV, 
and decreases to 6.5\% at 65 GeV. In the 1-tower data (23 GeV
photon trigger) the acceptance
increases from 35\% at 30 GeV to 43\% at 65 GeV. 

The efficiency of the remaining selection requirements are 
measured in the
data with a variety of complementary datasets \cite{Hall:2002sg}. 
The efficiency of the event $z_0$ requirement is measured to be 
$93.7 \pm 1.1$\% in minimum bias data. 
The conversion identification efficiency 
is measured using a loose sample of conversions
occurring in and around the CTC inner cylinder material.
We find an efficiency of $97.4 \pm 2.0$\% , 
where the uncertainty is determined by variations seen when 
dividing the data into $p_T$ bins . 
The 8 GeV electron
trigger efficiency is measured with a pre-scaled 5 GeV electron
trigger and an inclusive muon dataset, 
and has an asymptotic efficiency of $91.4 \pm 0.9$\%. 
The 23 GeV photon trigger efficiency is measured with pre-scaled
10 GeV and 23 GeV photon triggers, 
and has an efficiency of $91.4 \pm 4.3$\% \cite{Partos:2001qx}.
The electron identification efficiency is measured with the  
non-trigger electron in $Z \rightarrow e^+e^-$ data, and is found 
to be $84.3 \pm 3.0$\%.
The CTC tracking efficiency is measured with a track embedding
study, and has a plateau value of $96 \pm 2$\% per track above 400
MeV. The isolation
requirement efficiency is measured by choosing random locations in the
calorimeter in minimum bias data and adding up the energy found 
within a cone radius of 0.4. The efficiency is found to be 85.9 
$\pm$ 0.4 \%. 
The no-extra-track requirement efficiency is evaluated with electrons
in $Z \rightarrow
e^+e^-$ data, and is found to be 89.6 $\pm$ 0.5 \%. The missing energy
efficiency is evaluated with un-isolated conversion candidates.
These events are predominantly di-jet events where the true $\met$
is zero, so the measured $\met$ is due to the calorimeter 
resolution. The efficiency decreases from 1.0 at 20 GeV to 89\% at 65 GeV.

\begin{figure}
\scalebox{0.6}{\includegraphics{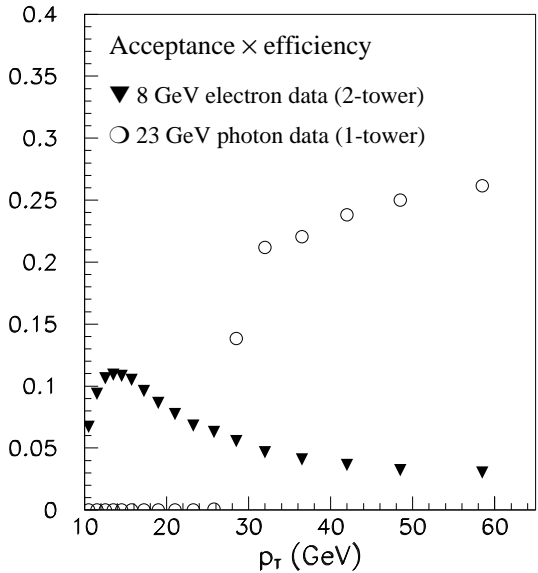}}
\caption{The total acceptance times efficiency for both 
conversion datasets. The decrease in the acceptance at high $p_T$ 
for the 8 GeV electron dataset is due to the 2-tower requirement, 
which becomes geometrically disfavored. The total conversion
probability is not included here.}
\label{fig:total_sig_eff}
\end{figure}

The efficiencies are summarized in Table \ref{tab:eff_summary}.
The total acceptance times efficiency for the two datasets is 
plotted in Figure \ref{fig:total_sig_eff}. 

\begin{table*}
\begin{center}
\begin{tabular}{|cccc|}
\hline
 source & 8 GeV electron  & 23 GeV photon & efficiency   \\
\hline
$z_0$ & * & * & $0.937 \pm 0.011$ \\
Conversion ID & * & * & $0.974 \pm 0.020$ \\
Level 1 trigger & * & * & 1.0 \\
Level 2 trigger (8 GeV) & * & & $91.4 \pm 0.9$ \% above 16 GeV \\
Level 2 trigger (23 GeV) & & * & $0.914 \pm 0.043$ \\
Level 3 electron ID & * & & $0.849 \pm 0.030$ \\
Tracking (CTC) & * & * & $0.96 \pm 0.02$ per track\\ 
Isolation & * & * & 0.859  $\pm$ 0.004 \\
No extra tracks & * & * & 0.896 $\pm$ 0.005 \\
$\met$ & * & & 1.0 below 20 GeV, 0.89 at 65 GeV  \\
\hline
\end{tabular}
\end{center}
\caption[] {Summary of signal efficiencies. The asterisks indicate
to which dataset each efficiency applies.}
\label{tab:eff_summary}
\end{table*}

%%%%%%%%%%%%%%%%%%%%%%%%%%%%%%%%%%%%%%%%%%%%%%%%%%%%%%%%%%%%%%%%%
%%%%%%%%%%%%%%%%%%%%%%%%%%%%%%%%%%%%%%%%%%%%%%%%%%%%%%%%%%%%%%%%

\section{Total conversion probability}
\label{sec:conv_prob}

The final element of the photon cross section measurement with 
conversions is the total probability that the photon converts in the 
CDF inner detector. The conversion probability has been evaluated
in several ways. The standard technique 
relies on a material map measured in the
data with an inclusive conversion dataset, calibrated with an 
\emph{a priori} determination of the amount of material 
in the CTC inner cylinder. 
This method gives a conversion probability of $5.17 \pm 0.28$\%, 
and we refer to this result as the standard material scale. 
Other material measurements based on the hard bremsstrahlung rate 
in $W \rightarrow e \nu$ and $J / \psi \rightarrow e^+e^-$ events 
give results in agreement with this number \cite{Affolder:2000bp}.

A second technique compares the number of $\pi^0$ Dalitz decays
($\pi^0 \rightarrow e^+e^- \gamma$) to the number of $\pi^0 \rightarrow
\gamma \gamma$ decays. In some Dalitz events
the on shell photon subsequently converts in the inner detector material.
Then the event has a four-track topology, with the invariant mass
being the $\pi^0$ mass. Similarly, in some $\gamma \gamma$ events both photons
convert in the detector material, giving the same four-track signature.
In the four-track event sample, the Dalitz events can be separated
from the $\gamma \gamma$ events because two of the Dalitz electrons
are prompt. Since the four-track Dalitz events undergo one conversion
in the detector material, while the four-track $\gamma \gamma$ events
undergo two conversions, the Dalitz-to-$\gamma \gamma$ ratio gives
the conversion probability, after accounting for the branching ratios
of the two decays.
This method gives a conversion probability of $8.02 \pm 0.73(stat)
 \pm 0.73(sys)$\%, which is significantly higher than the 
standard result quoted above.

Several other datasets also give evidence for a larger 
conversion probability. At CDF the reconstructed mass of di-muon 
resonances such as the $J / \psi$, $\psi(2s)$, $\Upsilon(1s)$,$(2s)$,
and $(3s)$ depend
on the amount of material in the inner detector due to muon $dE/dx$
energy losses. We correct the muon momenta for the expected 
energy loss by assuming the standard material scale. However, 
after the correction the reconstructed masses are less than
the PDG masses for all five resonances. For the $J / \psi$ 
the mass shift is more than twenty times larger 
than the statistical error, while for the $\Upsilon (3s)$ the
shift is only 1.2 times the statistical error. The 
dominant systematic uncertainty
is due the fact that the  
reconstructed $J / \psi$ mass depends on the amount of material
the muons pass through \cite{Affolder:2000bp}. After adopting a
systematic uncertainty
to account for this, the measured value of the $J / \psi$
mass agrees with the PDG value within errors. These effects
do not prove that the standard material scale is too small, but
they are consistent with that hypothesis.

There is also some evidence from W electrons for a larger material scale. 
In Run 1b the peak of the $E/p $ distribution in W electrons in the 
data
is shifted to the right with respect to the  Monte Carlo simulation 
when assuming the standard material scale. Although this effect is 
not adequately understood, the Monte Carlo $E/p$ peak can be made
to agree with the data by increasing the material \cite{Affolder:2000bp}.

In summary, the evidence concerning the total conversion probability
is ambiguous. Rather than choose between two conflicting results, we 
adopt the approach of choosing a central value and
systematic uncertainty which encompasses all possibilities. This
value is $ 6.60 \pm 1.43$\% . We make one adjustment to the conversion
probability to account for the effective loss of material due to the 
requirement on the radial location of the conversion ($r_{cnv} > 2.0$ cm). 
The final value of the effective conversion probability is 
$6.40 \pm 1.43$\%. The uncertainty on the conversion probability dominates 
all other errors on the photon cross section. However, while the conversion
probability affects the overall normalization of the cross section, 
it does not affect the shape of the cross section as a function
of $p_T$. This is assured by the fact that the pair-production 
cross section does not vary significantly over the $p_T$ range considered
here.

\section{Systematic uncertainties}

In this section we briefly summarize the systematic uncertainties
on the photon cross section measurement with conversions. 
(A detailed discussion is given in Ref. \cite{Hall:2002sg}.)
In Section \ref{sec:back_sub} 
we discussed the systematic uncertainties we adopt
to account for shortcomings in the Monte Carlo $E/p$ model, and
backgrounds due to prompt electrons. Both of these uncertainties
depend on $p_T$. 

For the 8 GeV electron data the remaining $p_T$ dependent systematic 
uncertainties are as follows. We take a 
systematic uncertainty to account for a possible time dependence on 
the trigger efficiency. This uncertainty is determined by counting the
fraction of events in our final sample which occur before the midpoint 
of Run 1b. This fraction is 53.6\%, and depends on $p_T$. We take 3.6\%
as the uncertainty. Secondly, the conversion identification efficiency
varies by 2.0\% when dividing the data into $p_T$ bins, and we take
this as a systematic uncertainty.

We also adopt the following $p_T$ independent systematic uncertainties.
The cross section uncertainty  
due the total conversion probability is +27/-17\%.
The CEM energy scale uncertainty results in an 
cross section error of 3.0\%. The integrated luminosity
is measured to 4.1\%, and the asymptotic trigger efficiency
is known to 1.4\%. There are also uncertainties due to the  
tracking efficiency (2.0\%),
the electron identification efficiency (3.5\%), 
and the $z_0$ requirement efficiency (1.2\%).

The systematic uncertainties on the 23 GeV photon data sample are similar, 
except there is no trigger time dependence, no electron identification
uncertainty, and the asymptotic trigger efficiency is known to 4.7\%. 

The total $p_T$ independent systematic uncertainty is +28/-18\% for both 
datasets. The total $p_T$ dependent systematic uncertainty in each $p_T$ 
bin is listed in Table \ref{tab:pt_dep_sys_err}. The total systematic 
uncertainty is listed in Table \ref{tab:cross_section_summary}.

\begin{table}[t!]
\begin{center}
\begin{tabular}{|cc|}
\hline
$p_T$ (GeV) & $p_T$ dep. sys. err. (\%)  \\
\hline
\multicolumn{2}{|c|}{ \bf{ 8 GeV electron (2-tower) data:}} \\
10-11 & +10.6/-12.8 \\
11-12 & +9.3/-11.6 \\
12-13 & +9.4/-9.3 \\ 
13-14 & +8.5/-8.6 \\
14-15 & +6.7/-7.3 \\
15-16.5 & +6.7/-6.9 \\
16.5-18 & +5.7/-6.0 \\
18-20 &  +7.6/-7.8 \\
20-22 &  +7.0/-6.1 \\
22-24.5 & +4.3/-5.8 \\
24.5-27 & +5.1/-11.9 \\
27-30 & +5.7/-11.3 \\
30-34 & +4.1/- 11.1\\
34-39 &  +4.1/-11.0 \\
39-45 & +5.6/-11.5 \\
45-52 & +4.1/-10.8 \\
52-65 & +8.8/-13.3 \\
\hline
\multicolumn{2}{|c|}{ \bf{ 23 GeV photon (1-tower) data:}} \\
30-34 & +2.3/-4.9 \\
34-39 & +2.8/-4.9 \\
39-45 &  +3.9/-5.6 \\
45-52 & +5.0/-4.7 \\
52-65 & +4.7/-8.2 \\ 
\hline

\end{tabular}
\end{center}
\caption{The $p_T$ dependent systematic uncertainty for all $p_T$ bins.
The correlated systematic uncertainty is +28/-18\% for both datasets.}
\label{tab:pt_dep_sys_err}
\end{table}

\section{Cross section measurement}

The cross section is calculated according to

\begin{equation}
        \frac{d\sigma^2}{d p_T d \eta} = 
	\frac{N_{signal}}{A \cdot \epsilon \cdot \Delta p_T
        \cdot\Delta \eta \cdot \int \mathcal{L}}
\end{equation}

$A \cdot \epsilon$
is the acceptance times efficiency shown in Figure \ref{fig:total_sig_eff}
multiplied by the effective conversion probability of 6.40\%.
We measure
the average cross section between -0.9 $ < \eta < $ 0.9, so
$\Delta \eta$ is 1.8. $\Delta p_T$ is the bin width,  
and $\int \mathcal{L}$ is the integrated luminosity of 73.6 pb$^{-1}$
for the 8 GeV electron data and 83.7 pb$^{-1}$ for the 23 GeV 
photon data.

The final result for both datasets is listed in Table
\ref{tab:cross_section_summary}.
In the $p_T$ region where the datasets overlap (30 GeV $< p_T <$ 65 GeV) 
the two measurements are in good agreement with each other.
This comparison is an important cross check on the acceptance and efficiency
calculations of the two datasets, since they differ by up to a factor 
of nine.

\begin{table*}
\begin{center}
\begin{tabular}{|cccccccc|}
\hline
$p_T$ & $ \langle p_T \rangle $  & & & $d \sigma / d p_T d\eta$ & stat & sys & NLO QCD \\
(GeV) & (GeV)  &  $\mathcal{A} \cdot \epsilon$ & $N_{signal}$ & (pb/GeV) & error (\%) & error (\%) & (pb/GeV) \\
\hline
\multicolumn{8}{|c|}{ \bf{ 8 GeV electron (2-tower) data:}} \\
10-11 & 10.5 & 0.067 & 7152 & 12590 & 2.2 & +30/-22 & 10968 \\
11-12 & 11.5 & 0.094 & 7761 & 9771 & 2.1 & +29/-22 & 7434 \\
12-13 & 12.5 & 0.106 & 6111 & 6773 & 2.2 & +29/-20 & 5203 \\ 
13-14 & 13.5 & 0.109 & 4320 & 4659 & 2.6 & +29/-20 & 3743 \\
14-15 & 14.5 & 0.108 & 3195 & 3483 & 2.9 & +29/-20 & 2758 \\
15-16.5 & 15.7 & 0.105 & 3059 & 2289 & 2.8 & +29/-20 & 1963 \\
16.5-18 & 17.2 & 0.096 & 1846 & 1509 & 3.5 & +28/-19 & 1328 \\
18-20 & 18.9 & 0.086 & 1391 & 950 & 4.1 & +29/-20 & 888 \\
20-22 & 20.9 & 0.077 & 863 &  658 & 5.1 & +29/-19 & 577 \\
22-24.5 & 23.2 & 0.068 & 596 & 413 & 6.0 & +28/-19 & 369 \\
24.5-27 & 25.7 & 0.063 & 344 & 258 & 7.7 & +28/-22 & 238 \\
27-30 & 28.3 & 0.056 & 272 & 207 & 8.8 & +28/-21 & 158 \\
30-34 & 31.9 & 0.047 & 136 & 85.9 & 13.5 & +28/-21 & 94.6 \\
34-39 & 36.3 & 0.041 & 101 & 58.5 & 14.4 & +28/-21 & 49.1 \\
39-45 & 41.6 & 0.036 & 63.9 & 34.5 & 18.7 & +28/-22 & 26.2 \\
45-52 & 48.1 & 0.032 & 21.7 & 11.4 & 53.3 & +28/-21 & 13.4 \\
52-65 & 57.8 & 0.030 & 16.6 & 5.0 & 33.5 & +29/-23 & 5.7 \\
\hline
\multicolumn{8}{|c|}{ \bf{ 23 GeV photon (1-tower) data:}} \\
30-34 & 31.9 & 0.212 & 723 & 88.4 & 4.8 & +28/-19 & 94.6 \\
34-39 & 36.3 & 0.220 & 564 & 53.1 & 5.3 & +28/-19 & 49.1 \\
39-45 & 41.6 & 0.238 & 316 & 22.9 & 7.4 & +28/-20 & 26.2 \\
45-52 & 48.1 & 0.250 & 225 & 13.3 & 8.5 & +29/-19 & 13.4 \\
52-65 & 57.8 & 0.261 & 131 & 4.0 & 11.3 & +29/-21 & 5.7 \\ 
\hline

\end{tabular}
\end{center}
\caption[8 GeV conversion cross section summary]
{Summary of the conversion cross section measurement in both datasets.
The 8 GeV electron data has an integrated luminosity of 
73.6pb$^{-1}$, and the 23 GeV photon data has an integrated luminosity
of 83.7pb$^{-1}$. 
$\Delta \eta$ is 1.8, and the effective conversion probability, 
which is not included in the acceptance $\times$ efficiency shown here, 
is 6.40\%. The NLO QCD theory was calculated by the authors of Reference 
\cite{Gluck:1994iz}, and uses the CTEQ5M parton distribution functions
with all scales set to the $p_T$ of the photon.}
\label{tab:cross_section_summary}
\end{table*}

\begin{figure}
\scalebox{0.6}{\includegraphics{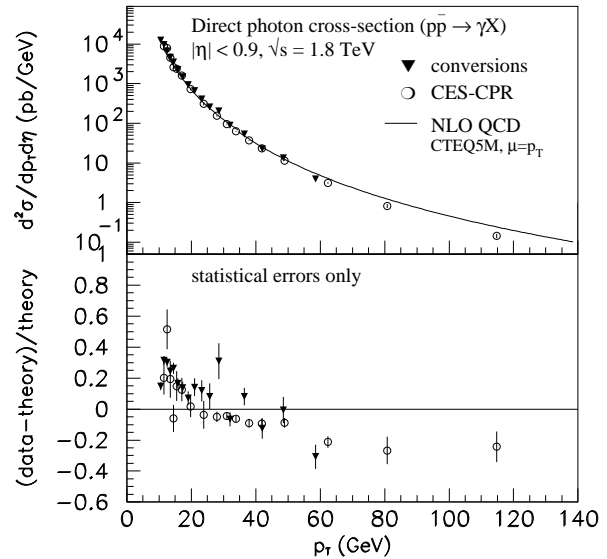}}
\caption[The isolated direct photon cross section]
{ The isolated direct photon cross section. 
The result of the conversion technique is compared with 
CES-CPR and theory. For the conversion measurement the
8 GeV electron data is shown below 30 GeV, and the 23 GeV
photon data above. The theory curve is from the authors of 
Reference  \cite{Gluck:1994iz}, and
uses the CTEQ5M parton distribution functions with the all
scales set to the $p_T$ of the photon. Only the statistical
error bars are shown here. 
% The CES-CPR measurement has a systematic 
%uncertainty of $\pm$10\%, while the conversion measurement has 
%a systematic uncertainty of +30\%/-20\%. For both measurements
%the systematic uncertainties affect primarily the overall normalization
%of the result.
}
\label{fig:xs_cnv_cescpr_qcd}
\end{figure}

Since the two datasets are in agreement we can combine the
measurements in the region of overlap. 
However, the 23 GeV photon
data would dominate the combined cross section (due to much
smaller errors), so instead we simply adopt 
the 23 GeV photon data above 30 GeV. This hybrid
cross section is compared to NLO QCD and the
standard CDF measurement (referred to as CES-CPR) in Figure 
\ref{fig:xs_cnv_cescpr_qcd}.
The theory curve is taken from the authors of Reference  \cite{Gluck:1994iz}.
The calculation uses the CTEQ5M parton distribution functions, 
and the renormalization, factorization, and fragmentation scales
have been set to the $p_T$ of the photon. This calculation takes
into account the suppression of the bremsstrahlung diagrams
due to the isolation requirement on the photon. In the lower half of Figure
\ref{fig:xs_cnv_cescpr_qcd} the measurements are shown as 
(data-theory)/theory.

%%%%%%%%%%%%%%%%%%%%%%%%%%%%%%%%%%%%%%%%%%%%%%%%%%%%%%%%%%%%%%%%%
%%%%%%%%%%%%%%%%%%%%%%%%%%%%%%%%%%%%%%%%%%%%%%%%%%%%%%%%%%%%%%%%

The CES-CPR measurement and the conversion measurement agree
with each other both in shape and in normalization. 
The total systematic uncertainty on the conversion measurement is larger
(+30\%/-20\%) than the CES-CPR measurement (18\% at 10 GeV 
and 11\% at 115 GeV) due to the large uncertainty on the total conversion
probability. Nevertheless, for both measurements the total 
systematic uncertainties are primarily $p_T$ independent,
so that both
techniques give a much more precise measurement of the shape of the
cross section as a function of $p_T$. 
The agreement of the conversion and CES-CPR measurements on the
shape is remarkable, since the two
techniques have  little in common with each other.
They use independent data samples, independent background subtraction 
techniques, and have  different acceptances, efficiencies, 
and systematic uncertainties. 

\begin{figure}
\scalebox{0.6}{\includegraphics{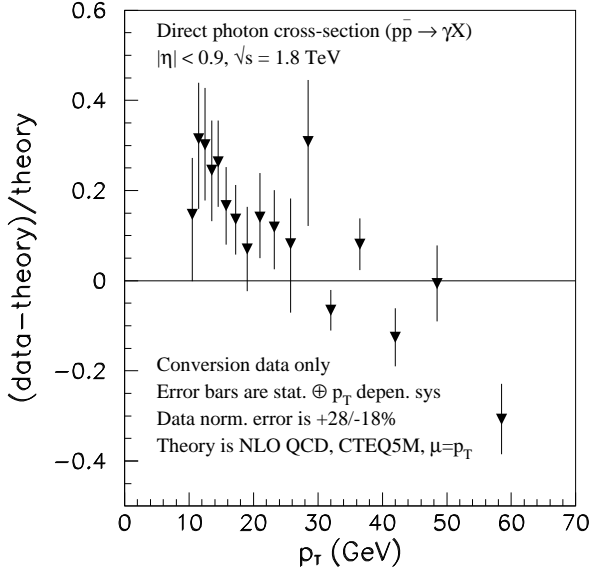}}
\caption[]
{The isolated photon cross section from conversions compared
to NLO QCD. The error bars shown here are the combined statistical
and $p_T$ dependent systematics, in order to compare the shape
of the measured cross section to theory.}
\label{fig:shape_analysis}
\end{figure}

Figure \ref{fig:shape_analysis} shows the conversion measurement 
alone as (data-theory)/theory. To compare the shape of the data 
to the calculation, the uncertainty bars in this plot are the 
combined statistical and $p_T$ dependent systematic uncertainties. The data
show a steeper slope than the calculation which is 
unexplained by the systematic uncertainties of the measurement. 
Other analyses have concluded that this type of shape difference
is difficult to resolve simply by changing the the renormalization, 
fragmentation, and factorization scales of the calculation,
or the set of parton distribution functions \cite{Acosta:2002ya}. 
Since two independent experimental techniques 
are in agreement on the shape, this is further evidence 
that refinements to the calculation are needed before
these measurements can provide useful constraints on the gluon
distribution of the proton.

%%%%%%%%%%%%%%%%%%%%%%%%%%%%%%%%%%%%%%%%%%%%%%%%%%%%%%%%%%%%%%%%%
%%%%%%%%%%%%%%%%%%%%%%%%%%%%%%%%%%%%%%%%%%%%%%%%%%%%%%%%%%%%%%%%

\section{Acknowledgements}
We would like to thank Werner Vogelsang for providing the theoretical calculations used in this paper. We also thank the Fermilab staff and the technical staffs of the participating institutions for their vital contributions. This work was supported by the U.S. Department of Energy and National Science Foundation; the Italian Istituto Nazionale di Fisica Nucleare; the Ministry of Education, Culture, Sports, Science and Technology of Japan; the Natural Sciences and Engineering Research Council of Canada; the National Science Council of the Republic of China; the Swiss National Science Foundation; the A.P. Sloan Foundation; the Bundesministerium fuer Bildung und Forschung, Germany; the Korean Science and Engineering Foundation and the Korean Research Foundation; the Particle Physics and Astronomy Research Council and the Royal Society, UK; the Russian Foundation for Basic Research; the Comision Interministerial de Ciencia y Tecnologia, Spain; work supported in part by the European Community's Human Potential Programme under contract HPRN-CT-20002, Probe for New Physics; and this work was supported by Research Fund of Istanbul University Project No. 1755/21122001.

\bibliography{myprd}% Produces the bibliography via BibTeX.

\end{document}